# Radiation of a bunch in a waveguide having areas with deeply corrugated and smooth walls

**Tatiana Yu. Alekhina and Andrey V. Tyukhtin**

Saint Petersburg State University
7/9 Universitetskaya nab., St. Petersburg, 199034 Russia

E-mails: t.alehina@spbu.ru, a.tyuhtin@spbu.ru

**Abstract**. We analyze radiation from a bunch that uniformly moves in a circular waveguide and crosses the boundary between two areas: with a corrugated wall and with a smooth wall. It is assumed that the corrugation period is much less than the wavelengths under consideration. However, in difference with our pervious works, the corrugation depth is of the same order as the wavelength. The influence of corrugation on the electromagnetic field is investigated using the method of equivalent boundary conditions. At first, we consider a regular corrugated waveguide. Then two cases of the waveguides with boundaries between the corrugated and smooth regions is studied: first, when the bunch flies out of the corrugated part into the smooth one, and second, when the bunch flies into the corrugated part from the smooth one. We obtain and analyze the analytical expressions for the field components in both areas and reveal the main physical peculiarities of the bunch radiation.

## 1. Introduction

Periodic waveguide structures have been increasingly attracting interest for their applications in accelerator and beam physics, as well as in microwave electronics since the 1970s [1,2]. Smith-Purcell radiation when the wavelengths are comparable to the structure period was considered (see, e.g. [3]). Recently, essential attention has been paid to the use of corrugated waveguides as sources of terahertz radiation [4–10].

Note that, along with Smith-Purcell radiation, a radiation at lower frequencies (when wavelengths are much more than the corrugation period) is also of interest. This radiation is due to the anisotropic nature of the equivalent boundary conditions (EBC), which describe periodic structures at such frequencies [11]. Such "longwave" problems was analytically analysed by using the EBC method for infinite plane structures [12-14] and for infinite waveguide [10, 15]. According to this approach, a corrugated structure is replaced with a smooth surface on which the EBC are fulfilled [11]. Note that the EBC have an analytical form, in particular, for the rectangular corrugation with a shallow profile and a deeply one (when the depth of the corrugation is of the same order as the wavelength) [11]. It was shown that the deep corrugation has essential advantages over the shallow one and can be useful, in particular, for bunch diagnostics.

Note that the case of waveguide where the charged particle bunch intersects the boundary between the rectangular corrugated waveguide area and the area with smooth wall was considered in [16] for the case of shallow profile. Here we analyze analogues problem but for the case of deep corrugation. Two cases are under investigation: first, when the bunch flies out of the corrugated part of the waveguide into the smooth one, and second, when the bunch flies into the corrugated part from the smooth one. The depth of the corrugation is supposed to be of the same order as the wavelength under consideration. The analytical solutions of the problems in both cases are obtained with the EBC method [11]. We will show that radiation can be both multimode and single mode depending on the problem parameters. We will reveal the main physical peculiarities of the bunch radiation in both cases under consideration.

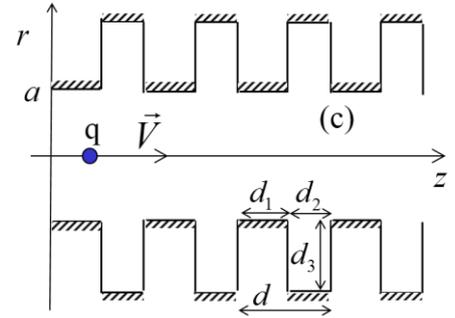

**Figure 1.** (Colour online) Geometry of the corrugated part of the waveguide in rz-section.

## 2. The problem statement and equivalent boundary conditions

We consider a bunch with a charge $q$ that moves with a constant velocity $\vec{V} = c\beta \vec{e}_z$ ($c$ is the light velocity in vacuum) in a metal circular waveguide of a radius $a$. Filling of the waveguide has properties of a vacuum. The bunch moves along the $z$ - axis of the cylindrical frame of reference $r$, $\varphi$, $z$. It is characterized by some distribution along the $z$ - axis and a negligible thickness, i. e., the charge density is

$$\rho = q\delta(r)\eta(\zeta)(2\pi r)^{-1}, \quad \text{where} \quad \int_{-\infty}^{\infty} \eta(\zeta)d\zeta = 1,$$

$\zeta = z - c\beta t$. For example, for the bunch with the Gaussian distribution

$$\eta(\zeta) = \exp\left(-\zeta^2 \left(2\sigma^2\right)^{-1}\right)\left(\sqrt{2\pi}\sigma\right)^{-1},$$

where σ is the half length of the bunch. The waveguide consists of two parts: the one part has a corrugated wall (c) and another has a smooth wall (s).

In the corrugated area (c), we consider the case of a deep corrugation, i.e. it is assumed that the depth of the structure $d_3$ is of the same order as the wavelength (or greater than it):

$$k_0 d_3 \geq 1, \quad (1)$$

where $k_0 = \omega/c = 2\pi/\lambda$ is a wave number. However, the structure period $d$ is much less than wavelength under consideration $\lambda$:

$$k_0 d \ll 1. \quad (2)$$

In this case, the corrugated structure can be replaced with a smooth surface on which so-called equivalent boundary conditions (EBC) are fulfilled [11]. The geometry of the corrugated part of the waveguide is shown in Fig. 1.

For the case under consideration, the EBC have the following form for the Fourier-transforms of electric and magnetic fields [11] in area (1):

$$E_{1z}\big|_{r=a} = g\, H_{1\varphi}\big|_{r=a}, \quad E_{1\varphi}\big|_{r=a} = 0. \quad (3)$$

Here, $g$ is impedance that is given by the expression

$$g = i\frac{d_2}{d}\frac{tg(k_0 d_3)}{1 - k_0 d\, l\, tg(k_0 d_3)}, \quad (4)$$

where $d_2$ is the width of the structure grooves. Parameter $l$ is determined by the formula

$$l = \frac{1}{2\pi}\left\{(2-\xi)^2 \ln(2-\xi) - \xi^2 \ln\xi - 2(1-\xi)\ln[4(2-\xi)]\right\}, \quad (5)$$

where $\xi = d_1/d$, $d_1 = d - d_2$. Dependence of this parameter on $\xi$ is shown in Fig. 2. As we can see it is positive and quite small: $0 \leq l < 0.082$. It should be also noted that EBC (4) can not be reduced to the EBC for shallow corrugation (where $d_3/d \ll 1$) [11] because this requires taking into account the terms of the order of $d_3/d$.

The bunch crosses the boundary between the corrugated and smooth regions (at $z=0$). Next, two cases are under investigation: first, when the bunch flies out of the corrugated part (c) of the waveguide into the smooth one (s), and second, when the bunch flies into the corrugated part (c) from the smooth one (s). The middle of the bunch intersects the point $z=0$ at the moment $t=0$.

The problem is formulated for the Fourier components of the magnetic field that must satisfy the equations

$$L_r H_\varphi^{(c,s)} + \frac{\partial^2}{\partial z^2} H_\varphi^{(c,s)} + k_0^2 \varepsilon\mu H_\varphi^{(c,s)} = \frac{4\pi}{c}\frac{\partial}{\partial r} j_\omega, \quad (6)$$

and the conditions

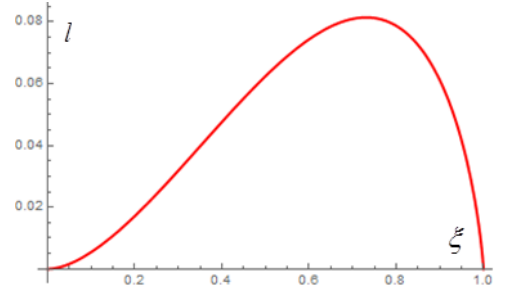

**Figure 2.** Dependence $l(\xi)$.

$$gH_\varphi^{(c)}\big|_{r=a} = \frac{i}{k_0 r}\frac{\partial}{\partial r}\left(rH_\varphi^{(c)}\right)\big|_{r=a}, \quad \frac{1}{r}\frac{\partial}{\partial r}\left(rH_\varphi^{(s)}\right)\big|_{r=a} = 0, \quad (7)$$

where

$$L_r = \frac{\partial^2}{\partial r^2} + \frac{1}{r}\frac{\partial}{\partial r} - \frac{1}{r^2},$$

$$j_\omega = \frac{q}{(2\pi)^2}\frac{\delta(r)}{r}\exp\left(\frac{i\omega z}{\beta c}\right)\tilde\eta(\omega).$$

Here $\tilde\eta(\omega)$ is the Fourier transform of the longitudinal distribution $\eta(\zeta)$, normalized to the distribution of a point charge. In particular,

$$\tilde\eta(\omega) = \exp\left[-\omega^2\sigma^2\left(2\beta^2 c^2\right)^{-1}\right]$$

for the bunch with the Gaussian distribution.

The components of the electric field are found in the form

$$E_r^{(c,s)} = -\frac{i}{k_0}\frac{\partial}{\partial z}H_\varphi^{(c,s)}, \quad E_z^{(c,s)} = \frac{i}{k_0}\left[\frac{1}{r}\frac{\partial}{\partial r}\left(rH_\varphi^{(c,s)}\right) - \frac{4\pi}{c}j_\omega\right],$$

$$E_\varphi^{(c,s)} = H_r^{(c,s)} = H_z^{(c,s)} = 0.$$

Therefore, we have TM-polarization only. Here, the field components for corrugated area (c) and smooth area (s) are labeled with superscript (c) and subscript (s), correspondingly.

The problem statement (6), (7) must be also supplemented with the matching conditions at $z=0$, which are the continuity conditions for tangential components of the total electromagnetic field. These conditions can be written in the form:

$$\frac{\partial}{\partial z}H_\varphi^{(c)}\big|_{z=0} = \frac{\partial}{\partial z}H_\varphi^{(s)}\big|_{z=0}, \quad H_\varphi^{(c)}\big|_{z=0} = H_\varphi^{(s)}\big|_{z=0}. \quad (8)$$

The field components in every area can be written in the form

$$\vec H^{(c,s)} = \vec H^{(c,s)q} + \vec H^{(c,s)b}, \quad \vec E^{(c,s)} = \vec E^{(c,s)q} + \vec E^{(c,s)b}. \quad (9)$$

The first summands in (9) (with superscript $q$) describe so-called 'forced' field, which is the field of a bunch in a regular infinite waveguide with properties of area (c) or (s). The second summands (with superscript $b$) are the 'free' field components that connected with the influence of the boundary between the corrugated and smooth areas of the waveguide (we follow the terminology from the book [18]).

For the Fourier transform of the forced electromagnetic field, the rigorous analytical solution of the problem (6), (7) is given in the form [16, 17]

$$H_{\varphi\omega}^{(c,s)q} = \frac{q|\omega|}{\pi c^2 \beta \gamma}\left[K_1(\kappa r) - R_{c,s} I_1(\kappa r)\right]e^{i\omega z/V}, \quad (10)$$

$$E_{r\omega}^{(c,s)q} = H_{\varphi\omega}^{(c,s)q}\beta^{-1},$$

$$E_{z\omega}^{(c,s)q} = -\frac{iq\omega}{\pi c^2 \beta^2 \gamma^2}\left[K_0(\kappa r) + R_{c,s} I_0(\kappa r)\right]e^{i\omega z/V},$$

$$R_c = -\frac{K_0(\kappa a) + \beta\gamma g_0 K_1(\kappa a)}{I_0(\kappa a) - \beta\gamma g_0 I_1(\kappa a)}, \quad R_s = -\frac{K_0(\kappa a)}{I_0(\kappa a)}, \quad (11)$$

where $g_0 = \operatorname{Im} g$ (4), $I_{0,1}(x)$ are modified Bessel functions of the first kind, $K_{0,1}(x)$ are Macdonald functions, $\gamma = (1-\beta^2)^{-1/2}$ is the Lorentz factor and $\kappa = |\omega|(c\beta\gamma)^{-1}$.

Note that for waveguide with smooth walls, the solution is well known and another method based on the spectral problem with eigenfunctions $J_1(\chi_n^{(s)} r)$ and eigenvalues

$$\chi_n^{(s)} = \chi_{0n}/a, \quad J_0(\chi_{0n}) = 0, \quad n=1,2\ldots \quad (12)$$

give expressions in the form of decomposition in an infinite series of normal modes. These functions have the following orthogonality property:

$$\int_0^a J_1(\chi_m^{(s)} r) J_1(\chi_n^{(s)} r) r dr = a^2 J_1^2(\chi_{0n})\delta_{mn}/2. \quad (13)$$

### 3. Radiation of the bunch in the regular deeply corrugated waveguide

Here, we analyze the Fourier integral for the forced field (10), (11) for an infinite waveguide with a corrugated wall (c). The analytical investigation of (10) is based on the complex variable function theory [20] as well as in our previous studies [21-23]. Further, we will note the main peculiarities of the forced field only. Investigation of the integrand (10) on the complex plane of $\omega$ shows that there are poles $\pm\omega_{0m}$ ($m=1,\ldots$) at the real axis, which give radiation with the discrete spectrum. Note that in this situation, there is no threshold value of the bunch velocity as opposed to the case of semi-infinite media with analogous corrugated walls [15]. The poles are obtain from the dispersion equation

$$F(x) \equiv I_0(x) - \beta\gamma g_0(x)I_1(x) = 0, \quad (14)$$

where $x = \kappa(\omega)a$ and

$$g_0(x) = \frac{d_2}{d}\frac{tg(\beta\gamma x d_3 a^{-1})}{1 - \beta\gamma x d a^{-1} l\, tg(\beta\gamma x d_3 a^{-1})}. \quad (15)$$

The solution of (14) $x_m = a\kappa(\omega_{0m})$ can be obtained numerically and analytically using iteration procedure [21] in non- and ultrarelativistic cases. So, one can obtain approximate analytic expressions for the dispersion equation solution $x_m$ in the case $|x| \gg 1$, which corresponds to the nonrelativistic case $\beta \ll 1$,

$$x_m \approx \frac{a\pi}{\beta d_3}\left(m - \frac{1}{2}\right)\left[1 - \frac{d \cdot l}{d_3} - \frac{d_2 \cdot l}{d_3}\frac{\beta d_3}{a\pi(m-1/2)}\right], m=1,\ldots (16)$$

and in the ultrarelativistic case $\gamma \gg 1$, or $|x| \ll 1$:

$$x_1 \approx \frac{a}{\gamma d_3}\frac{1}{\sqrt{s_0 + 1/3}}, \quad s_0 = \frac{d}{d_3}\left(l + \frac{d_2 a}{2d^2}\right), \quad (17)$$

$$x_m \approx \frac{a}{\gamma d_3}\left[(m-1)\pi + \frac{1}{s_0(m-1)\pi}\right], \quad m=2,\ldots \quad (18)$$

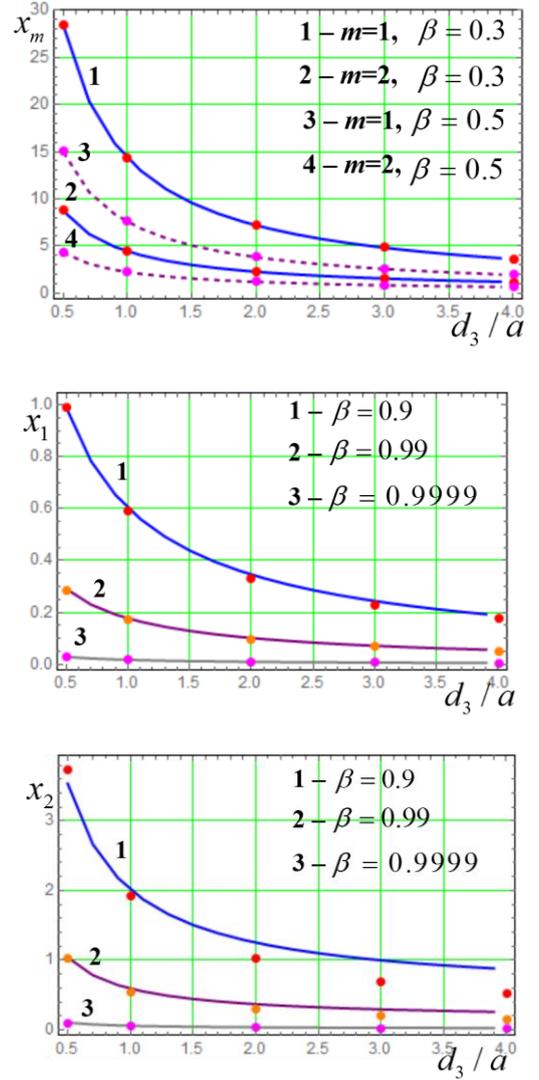

**Figure 3.** The dependence of the dispersion equation solution on the corrugation depth $d_3$ for different mode numbers m and velocities $\beta$: $x_m$ (16) – (top), $x_1$ (17) - (middle), and $x_2$ (18) - (bottom). Results of numerical solution $x_m$ are shown with circles. The waveguide parameters are the following: $a = 1$ cm, $d = 0.13$ cm, $d_1 = 0.05$ cm.

Figure 3 show the relevance of Exp. (16) – (18) to values $x_m$ computed from (14) for different parameters of the problem. Figure 3 (top) and (middle and bottom) correspond to the relatively small velocities and ultrarelativistic case, respectively. We can see a very good agreement when the usability condition is fulfilled.

Integrating (10) with Jordan's lemma and taking into account the poles $\pm\omega_{0m}$ only, we obtain the forced field components with discrete spectrum in the form:

$$\begin{Bmatrix} H_{1\varphi}^{(c)qd} \\ E_{1r}^{(c)qd} \end{Bmatrix} = \frac{q}{a^2}\sum_{m=1}^{\infty}\begin{Bmatrix} AH_{\varphi m}^{(c)qd} \\ AE_{rm}^{(c)qd} \end{Bmatrix}\mathrm{Sin}\left[\frac{\omega_{0m}}{c\beta}\zeta\right]\theta(-\zeta),$$

$$E_{1z}^{(c)qd} = \frac{q}{a^2}\sum_{m=1}^{\infty}AE_{zm}^{(c)qd}\mathrm{Cos}\left[\frac{\omega_{0m}}{c\beta}\zeta\right]\theta(-\zeta),$$

$$AH_{\varphi m}^{(c)qd} = -4\beta\gamma\tilde{\eta}(\omega_{0m})G_m^{-1}I_1\left(\frac{x_m r}{a}\right),\ AE_{rm}^{(c)qd} = \beta^{-1}AH_{\varphi m}^{(c)qd},$$

$$AE_{zm}^{(c)qd} = 4G_m^{-1}\tilde{\eta}(\omega_{0m})I_0\left(\frac{x_m r}{a}\right),\quad (19)$$

where

$$G_m = I_1(x_m)\frac{d}{dx}F(x)\bigg|_{x=x_m},\quad (20)$$

$F(x(\omega))$ is described by (14) and $\theta(x)$ is the Heaviside step function.

One can obtain analytical expressions for the amplitudes $AH_{\varphi m}^{qd}$, $AE_{rm}^{qd}$, $AE_{zm}^{qd}$ in different cases:

- in the nonrelativistic case $\beta \ll 1$ (or $|x| \gg 1$)

$$G_m^{-1} \approx -\frac{2d_2 a^2}{l^2 d^3 \beta(m-1/2)}\left[1 - \frac{d\cdot l}{d_3} - \frac{d_2\cdot l\beta}{a\pi(m-1/2)}\right]^{-1}e^{-2x_m},\ (21)$$

where $x_m$ is described with (16);

- in the ultrarelativistic case $\gamma \gg 1$ (or $|x| \ll 1$):

$$G_m^{-1} = -\left[\frac{1}{2} + \frac{d(d_3 + ld)}{ad_2}\right]^{-1}\left[1 + O(x_m^2)\right],\quad (22)$$

| $m$ | $x_m$ | $f_{0m}$, GHz | $AE_{zm}^{(c)qd}$ | Amplitude $E_{z1}^{(c)qd}$ on axis (KV/m) |
|---|---|---|---|---|
| | | $d_3 = 0.5$ cm | | |
| 1 | 0.029 | 9.8 | 0.498 | 192 |
| 2 | 0.101 | 34.1 | 0.037 | 14 |
| 3 | 0.185 | 62.5 | 0.003 | 0.12 |
| | | $d_3 = 1$ cm | | |
| 1 | 0.017 | 5.7 | 0.419 | 150 |
| 2 | 0.055 | 18.2 | 0.146 | 53 |
| 3 | 0.096 | 32.1 | 0.026 | 9.5 |

**Table 1.** Analytic calculation for the forced field in corrugated area for the following problem parameters: $a = 1$ cm, $d = 0.13$ cm, $d_1 = 0.05$ cm, $r = 0$, $\sigma = 0.25$ cm for the bunch with the Gaussian distribution, $q = -1$ nC, and $\beta = 0.9999$ (the Lorentz factor $\gamma = 70.7$).

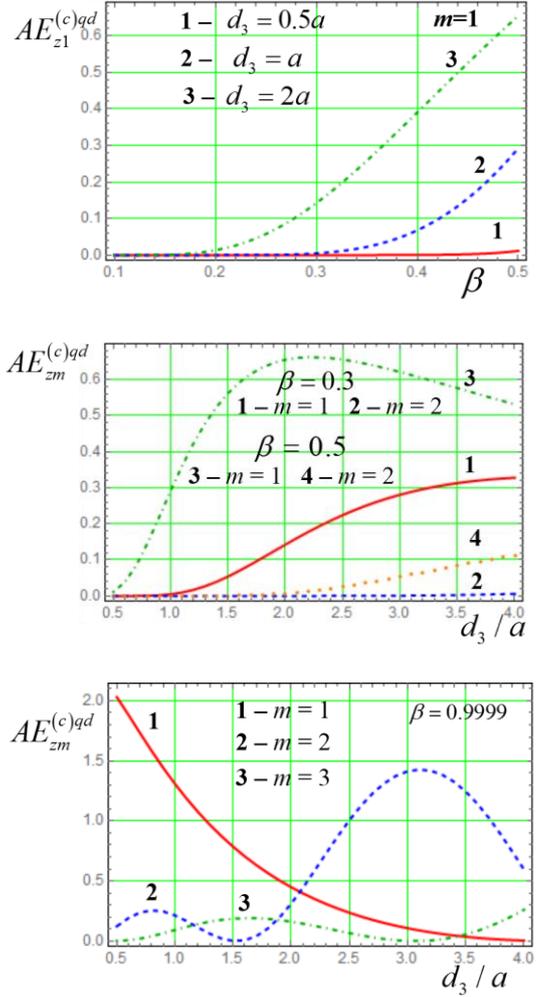

**Figure 4.** The behaviour of the amplitude $AE_{zm}^{(c)qd}$ (19) at different parameters of the problem for the bunch with the Gaussian distribution in the nonrelativistic case (top and middle) and in the ultrarelativistic case (bottom) at different corrugation depths $d_3$, mode numbers m and different velocities of the bunch motion $\beta$. The bunch and the waveguide parameters are the following: $\sigma = 0.25$ cm, $r = 0$, $a = 1$ cm, $d = 0.13$ cm, $d_1 = 0.05$ cm.

where $x_1$ and $x_m, m = 2,...$ are described with (17) and (18). Note that, in the ultrarelativistic case, the amplitude does not depend on the velocity of the bunch motion and it is approximately constant (22) that only depends on the waveguide parameters.

The behavior of the amplitude $AE_{zm}^{(c)qd}$ at different corrugation depths $d_3$, mode numbers m and velocities of the bunch motion $\beta$ for the bunch with the Gaussian distribution is shown in Fig. 4. The examples of calculations according to the formulae (14) are also presented in Table 1. One can see that, in the ultrarelativistic case, the amplitude of the first mode for the case of $d_3 = 0.5$ cm is much more than the amplitudes of other modes, so quasi-monochromatic radiation is generated for the problem parameters under consideration. In the case of $d_3 = 1 \text{cm}$, two modes play the main role in the radiation field.

## 4. The case of flying of the bunch from the corrugated area into the smooth one

### 4.1. General solution

Let us consider the case when the charge flies from the corrugated area into the smooth one. General solution of the problem under consideration is found in the form (9) where the forced fields in both areas are described by (10), (11). The free fields in the corrugated and smooth domains are presented in the form of decomposition in an infinite series of normal modes. For waveguide with smooth walls, eigenvalues $\chi_n^{(s)}$ are described by (12) and the eigenfunctions are $J_1\left(\chi_n^{(s)} r\right)$ with the orthogonality property (13). For waveguide with corrugated walls, the solution is based on the spectral problem for eigenfunctions $J_1\left(\chi_m^{(c)} r\right)$:

$$\begin{cases} L_r J_1\left(\chi^{(c)} r\right) = -\chi^{(c)2} J_1\left(\chi^{(c)} r\right) \\ -ik_0 g J_1\left(\chi^{(c)} r\right)\Big|_{r=a} = \frac{1}{r}\frac{\partial}{\partial r}\left(r J_1\left(\chi^{(c)} r\right)\right)\Big|_{r=a} \end{cases} \quad (23)$$

and eigenvalues $\chi_m^{(c)}$ are determined by the dispersion equation

$$\chi J_0(\chi a) - k_0 g_0 J_1(\chi a) = 0, \quad m = 1, 2, \ldots. \quad (24)$$

One can show that radiation with the discrete spectrum is generated with frequencies

$$\omega_{0m} = c\beta\gamma\left|\chi_m^{(c)}\right|, \quad (25)$$

which are real if the solution of the dispersion equation (24) is imaginary $\chi_m^{(c)} = i\left|\chi_m^{(c)}\right|$. Then, $k_{zm}^{(c)}$ is described by the formula $k_{zm}^{(c)} = \frac{1}{c}\sqrt{\omega^2 + \tilde{\Omega}_m^2}$, $\tilde{\Omega}_m = \left|\chi_m^{(c)}\right|c$.

So, the free field in the both area can be obtained in the form of decomposition in an infinite series of normal modes. For example, we only give expression for the $\varphi$-components of the magnetic field:

$$H_\varphi^{(c,s)b} = \frac{2q}{\pi c a^3}\sum_{m=1}^{\infty} J_1\left(\chi_m^{(c,s)} r\right) \times \\ \times \int_{-\infty}^{+\infty} B_m^{(c,s)} \tilde{\eta}(\omega)\exp\left[i\left(k_{zm}^{(c,s)}|z| - \omega t\right)\right]d\omega, \quad (26)$$

where

$$\Omega_m = \chi_m^{(c)} c, \quad k_{zm}^{(c)} = \frac{1}{c}\sqrt{\omega^2 - \Omega_m^2}, \\ \omega_n = \chi_n^{(s)} c = \chi_{0n} c/a, \quad k_{zn}^{(s)} = \frac{1}{c}\sqrt{\omega^2 - \omega_n^2}. \quad (27)$$

Here, $\text{Im}\, k_{zm}^{(c,s)} \geq 0$. This condition means that the waves outgoing from the boundary must decrease exponentially with an increase in the distance $|z|$ if dissipation in the medium is taken into account.

For coefficients $B_n^{(c,s)}$ we use the matching conditions (8) at $z = 0$. Then, we multiply the obtained expressions by $J_1(\chi_{0n} r/a)$, integrate and use the orthogonality condition (13) and known integral [19]:

$$\int_0^a r J_1(\chi r) J_1(\chi_{0n} r/a) dr = \frac{a^3 \chi}{\chi_{0n}^2 - a^2 \chi^2} J_0(\chi a) J_1(\chi_{0n}) \quad (28).$$

So, coefficients $B_n^{(s)}$ are equal to

$$B_n^{(s)} = U_n + \sum_{m=1}^{\infty} \alpha_{mn} B_m^{(c)}, \quad (29)$$

where

$$\alpha_{mn} = \frac{2a\chi_m^{(c)} J_0\left(\chi_m^{(c)} a\right)}{\left(\chi_{0n}^2 - a^2 \chi_m^{(c)2}\right) J_1(\chi_{0n})},$$

$$U_n = \frac{a^4 \kappa^2 (R_s - R_c) I_0(\kappa a)}{\left(\chi_{0n}^2 + a^2 \kappa^2\right) J_1(\chi_{0n})}$$

and coefficients $B_m^{(c)}$ are found from the equation systems

$$\sum_{n=1}^{\infty} M_{mn} B_n^{(c)} = U_m\left(k_{zm}^{(s)} - \frac{\omega}{c\beta}\right), \quad m = 1, 2\ldots, \quad (30)$$

where

$$M_{nm} = \alpha_{mn}\left(k_{zm}^{(s)} - k_{zn}^{(c)}\right). \quad (31)$$

### 4.2. Radiation of the bunch in the smooth area of the waveguide

Now, we analyze the free field (26). The main attention will be paid to the discrete spectrum of the radiation in the smooth area of the waveguide. We analyse Eqs. (26), (29) - (31) for coefficients $B_m^{(c,s)}$ using analytical methods based on the complex variable function theory [18]. Analogues studies were performed in our previous papers [22-24] for different problems with sectional homogeneous waveguides.

The investigation of the integrand singularities on the complex plane of $\omega$ for the modes with numbers m and n of the free field components show that along the poles $\pm\omega_{0m}$ (25), which give the discrete spectrum radiation, there also are the poles on the imaginary axis $\pm\omega_{0n}^{(s)} = \pm i\beta\gamma\omega_n$ and the branch points of the radicals $k_{zm}^{(c,s)}$ ($\pm i\tilde{\Omega}_m$ and $\pm\omega_n - i0$ accordingly).

Further, we will only consider radiation with the discrete spectrum. This is connected with the contribution of the poles (25) which are the solution of the dispersion equation (14) in the free field and determined by the integrals of view

$$I_n^{(s)b} = \int_{-\infty}^{+\infty} \frac{\tilde{B}_n^{(s)}(\omega)}{F(\omega)} \tilde{\eta}(\omega)\exp\left[i\left(k_{zn}^{(s)} z - \omega t\right)\right]d\omega, \quad (32)$$

where $\tilde{B}_n^{(s)}(\omega)$ is a smooth function of $\omega$ and does not have poles (25):

$$\tilde{B}_n^{(s)}(\omega) = V_n(\omega) + \sum_{m=1}^{\infty} \alpha_{mn}(\omega)\tilde{B}_m^{(c)}(\omega), \quad (33)$$

where

$$\chi_m^{(c)} = \chi_m^{(c)}(\omega), \; V_n(\omega) = U_n(\omega)F(\omega)$$

and coefficients $\tilde{B}_m^{(c)}(\omega)$ are found from the equation systems

$$\sum_{n=1} M_{mn}\tilde{B}_n^{(c)} = V_m\left(k_{zm}^{(s)} - \frac{\omega}{c\beta}\right), \quad m = 1, 2, \ldots \quad (34)$$

Asymptotic expression for integral (32) can be obtained with the steepest descent method (SDM) [18] as well as in our works [16, 22-24]. The integral behavior investigation by SDM includes saddle point determination, steepest descent path (SDP) building and transformation of the initial integration path toward the SDP. The poles (25) can be crossed during this transformation, and the contribution of the corresponding singularities can be include in the free field.

The pole contributions to integral (32) can be calculated using the residue theorem and we obtain the following expressions:

$$I_n^{(s)b} = 2\mathrm{Re}\left\{\begin{array}{l}\tilde{B}_n^{(s)}(\omega)\tilde{\eta}(\omega)\exp\left[i\left(k_{zn}^{(s)}z - \omega t\right)\right]\Big|_{\omega=\omega_{0m}} \times \\ \times \underset{\omega=\omega_{0m}}{\mathrm{Res}}\left[F^{-1}(\omega)\right]\end{array}\right\},$$

Analysis of the system (34) at $\omega = \omega_{0m}$ shows that this system has the following form

$$\sum_{n=1} M_{mn}(\omega_{0m})\tilde{B}_n^{(c)}(\omega_{0m}) = N_m(\omega_{0m}), \quad m = 1, 2\ldots, \quad (35)$$

and m-th column of $M_{mn}(\omega_{0m})$ has the form $M_{mm}(\omega_{0m}) = N_m(\omega_{0m})C_m(\omega_{0m})$. To derive the Eq. (35) we used the equality $\chi_m^{(c)}(\omega_{0m}) = i\kappa(\omega_{0m}) = x_m/a$ which results from coincidence between dispersion equation (14) and (24) at frequencies $\omega = \omega_{0m}$. Hence, the usage of the Cramers rule for solving the system (35) leads to the following result:

$$\tilde{B}_m^{(c)}(\omega_{0m}) = C_m^{-1}(\omega_{0m}), \; \tilde{B}_n^{(c)}(\omega_{0m}) = 0, \quad n \neq m \;, \quad (36)$$

$$C_m^{-1}(\omega_{0m}) = \frac{ia^2}{2I_1(x_m)}\left(k_{zm}^{(s)} - \frac{\omega}{c\beta}\right)\left(k_{zm}^{(s)} + \frac{\omega}{c\beta}\right)^{-1}\Bigg|_{\omega=\omega_{0m}}.$$

So, one can obtain the free field components with the discrete part of spectrum in the form

$$\begin{Bmatrix}H_\varphi^{(s)bd}\\E_r^{(s)bd}\end{Bmatrix} = \frac{q}{a^2}\sum_{n,m=1}^{\infty}\begin{Bmatrix}AH_{\varphi nm}^{(s)bd}\\AE_{rnm}^{(s)bd}\end{Bmatrix}\mathrm{Sin}\left[\frac{\omega_{0m}}{c\beta}(\Psi_{nm}z-t)\right]\theta\left(v_{gnm}^{(s)}t-z\right),$$

$$E_z^{(s)bd} = \frac{q}{a^2}\sum_{n,m=1}^{\infty} AE_{znm}^{(s)bd}\mathrm{Cos}\left[\frac{\omega_{0m}}{c\beta}(\Psi_{nm}z-t)\right]\theta\left(v_{gnm}^{(s)}t-z\right),$$

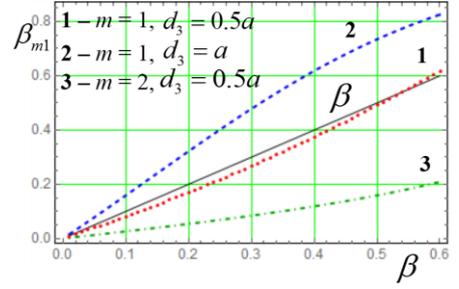

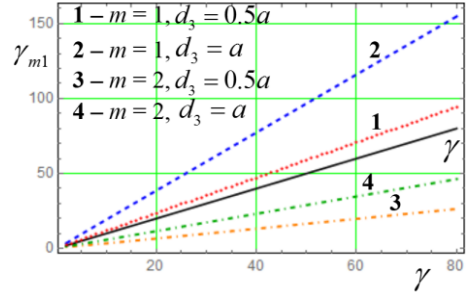

**Figure 5.** The threshold value $\beta_{mn}$ (and $\gamma_{mn}$) (38) as a function of the bunch velocity $\beta$ (or $\gamma$) at different corrugation depths $d_3$ and mode numbers $m$. The bunch and waveguide parameters are the same as in Fig. 4, $n = 1$.

$$AH_{\varphi nm}^{(s)bd} = \frac{8\beta\gamma I_1(x_m)}{G_m J_1^2(\chi_{0n})(\chi_{0n}^2 + x_m^2)}\frac{\beta\Psi_{mm}-1}{\beta\Psi_{mm}+1}J_1\left(\frac{\chi_{0n}r}{a}\right)\tilde{\eta}(\omega_{0m})$$

$$AE_{rnm}^{(s)bd} = \beta^{-1}AH_{\varphi nm}^{(s)bd},$$

$$AE_{znm}^{(s)bd} = \frac{8\chi_{0n}I_0(x_m)}{G_m J_1^2(\chi_{0n})(\chi_{0n}^2 + x_m^2)}\frac{\beta\Psi_{mm}-1}{\beta\Psi_{mm}+1}J_0\left(\frac{\chi_{0n}r}{a}\right)\tilde{\eta}(\omega_{0m}),$$

$$\Psi_{nm} = \sqrt{1-\chi_{0n}^2(\beta\gamma x_m)^{-2}}. \quad (37)$$

The mode with numbers m and n is a part of the wave field if

$$\beta > \beta_{mn} = \left(1+x_m^2\chi_{0n}^{-2}\right)^{-1/2} \text{ or } \gamma > \gamma_{mn} = \sqrt{1+\chi_{0n}^2 x_m^{-2}}. \quad (38)$$

If $\beta < \beta_{mn}$ (or the Lorentz factor $\gamma < \gamma_{mn}$) then the mode is totally reflected from the border between the corrugated and smooth parts of the waveguide. The threshold value $\beta = \beta_{mn}$ decreases with an increase in number m and increases with an increase in number n (the opposite takes place for the threshold value of the Lorentz factor $\gamma = \gamma_{mn}$). The behaviour of the threshold values $\beta_{mn}$ (or $\gamma_{mn}$) (38) at different bunch and waveguide parameters are presented in Fig. 5 for the nonrelativistic case (top) and in the ultrarelativistic case (bottom) when $x_m, m = 1, \ldots$ are described by (16) - (18). In the vacuum area, the discrete mode radiation exists in the domain

$$z < tv_{gnm}^{(s)}, \; v_{gnm}^{(s)}(\omega_{0m}) = c\Psi_{nm}, \quad (39)$$

where $\Psi_{nm}$ is described by (37) and $v_{gnm}^{(s)}$ is group velocity of the wave in the smooth area of the waveguide.

As one can see this radiation in the smooth area of the waveguide is always multi-mode with numbers m and n. However, analysis shows that the radiation generated in the corrugated area has appreciable amplitude for several first mode number m, so only several mode with number m can penetrate into the smooth area (depending on the condition (35)). Results of calculations are presented in Table 2 for the same parameters of the problems as in Table 1. In both cases under consideration the first modes are totally reflected and do not penetrate into the smooth part of the waveguide. Thus, it is possible to generate both multi-mode and quasi single-mode radiation in the smooth area depending on the parameters of the problem.

## 5. The case of flying of the bunch from the smooth area into the corrugated one

In this section, we consider the case when the bunch is flying out of the smooth area (s) at $z < 0$ into the corrugated one (c) at $z > 0$. We omit the description of the solution of the problem that is analogues to the first case (Section 4) and give the results only.

General expressions for the $\varphi$ – components of the magnetic fields in both area are the same as in the first case (26). Coefficients $B_n^{(c,s)}$ are obtained with (8), (13) and (28) as well as in the previous case. Coefficients $B_n^{(s)}$ are found from (29) and coefficients $B_m^{(c)}$ are the solution of the equation systems

$$\sum_{n=1} A_{nm} B_n^{(c)} = -U_m\left(k_{zm}^{(s)} + \frac{\omega}{c\beta}\right), \quad m = 1, 2\ldots, \quad (40)$$

where

$$A_{nm} = \alpha_{nm}\left(k_{zm}^{(s)} + k_{zn}^{(c)}\right). \quad (41)$$

Just as in the previous case, we will consider only the free field discrete part. This is connected with the contribution of the poles (25) in the free field that determined in the smooth and corrugated regions by the integrals of view

$$I_n^{(c,s)b} = \int_{-\infty}^{+\infty} \frac{\tilde{B}_n^{(c,s)}(\omega)}{F(\omega)} \tilde{\eta}(\omega) \exp\left[i\left(k_{zn}^{(c,s)} z - \omega t\right)\right] d\omega, \quad (42)$$

where coefficients $\tilde{B}_n^{(c,s)}(\omega)$ are smooth functions of $\omega$ and does not have poles (25). Coefficients $\tilde{B}_m^{(c)}(\omega)$ are found from the equation systems

$$\sum_{n=1} A_{mn} \tilde{B}_n^{(c)} = -V_m\left(k_{zm}^{(s)} + \frac{\omega}{c\beta}\right), \quad m = 1, 2\ldots. \quad (43)$$

The pole contributions to integrals (42) can be calculated using the residue theorem and we obtain the following expression

$$I_{mn}^{(c,s)b} = 2\text{Re}\left\{ \begin{array}{l} \tilde{B}_n^{(c,s)}(\omega)\tilde{\eta}(\omega)\exp\left[i\left(k_{zn}^{(c,s)} z - \omega t\right)\right]\Big|_{\omega=\omega_{0m}} \times \\ \times \underset{\omega=\omega_{0m}}{\text{Res}}\left[F^{-1}(\omega)\right] \end{array} \right\}, \quad (44)$$

| m | $f_{0m}$, GHz | $\gamma_{m1}$ | Amplitude $E_{zm}^{(c)qd}$ on axis (KV/m) | Amplitude $E_{zm}^{(s)bd}$ on axis (KV/m) |
|---|---|---|---|---|
| $d_3 = 0.5$ cm | | | | |
| 1 | 9.8 | 84.5 | 192 | - |
| 2 | 34.1 | 23.8 | 14 | 9 |
| 3 | 62.5 | 13.1 | 0.12 | 0.05 |
| $d_3 = 1$ cm | | | | |
| 1 | 5.7 | 158.1 | 150 | - |
| 2 | 18.2 | 43.8 | 53 | 20 |
| 3 | 32.1 | 25.0 | 9.5 | 1 |

**Table 2.** Analytic calculation for the free field for the same problem parameters as in Table 1.

Analysis of the system (43) at $\omega = \omega_{0m}$ shows that this system has the following form

$$\sum_{n=1} A_{nm}(\omega_{0m}) \tilde{B}_m^{(c)}(\omega_{0m}) = P_m(\omega_{0m}), \quad m = 1, 2\ldots, \quad (45)$$

and m-th column of $A_{nm}(\omega_{0m})$ has the form $A_{mm}(\omega_{0m}) = P_m(\omega_{0m}) Q_m(\omega_{0m})$. Hence, the usage of the Cramers rule for solving the system (35), as well as in the previous case, leads to the following result:

$$\tilde{B}_m^{(c)}(\omega_{0m}) = Q_m^{-1} = \frac{ia^2}{2I_1(x_m)}, \quad \tilde{B}_n^{(c)}(\omega_{0m}) = 0, \ n \neq m. \quad (46)$$

So, one can obtain that, in the smooth area of the waveguide,

$$\tilde{B}_n^{(s)}(\omega_{0m}) = 0, \quad n, m = 1, 2\ldots$$

and the discrete spectrum radiation is absent. In the corrugated area, the amplitude of every m-th mode of the free field discrete part is equal and opposite in sign to the amplitude of the forced discrete pare. Hence, there is the compensation effect in some domain near the boundary $z < v_{gm}^{(c)} t$, the group velocity $v_{gm}^{(c)}$ is described with

$$v_{gm}^{(c)} = c\left(\omega + \tilde{\Omega}_m \partial \tilde{\Omega}_m / \partial \omega\right)^{-1} \sqrt{\omega^2 + \tilde{\Omega}_m^2}\Big|_{\omega=\omega_{0m}}. \quad (47)$$

For the free field components with the discrete part of spectrum, we have:

$$\left\{ \begin{array}{l} H_\varphi^{(c)bd} \\ E_r^{(c)bd} \end{array} \right\} = -\frac{q}{a^2} \sum_{m=1}^{\infty} \left\{ \begin{array}{l} AH_{\varphi m}^{(c)qd} \\ AE_{rm}^{(c)qd} \end{array} \right\} \text{Sin}\left[\frac{\omega_{0m}}{c\beta}\zeta\right] \theta(-\zeta)\theta\left(v_{gm}^{(c)} t - z\right),$$

$$E_z^{(c)bd} = -\frac{q}{a^2} \sum_{m=1}^{\infty} AE_{zm}^{(c)qd} \text{Cos}\left[\frac{\omega_{0m}}{c\beta}\zeta\right] \theta(-)\theta\left(v_{gm}^{(c)} t - z\right),$$

where the amplitudes $AH_{\varphi m}^{qd}$ $AE_{rm}^{qd}$, $AE_{zm}^{qd}$ are described by (19).

## 6. Conclusion

The electromagnetic field of a bunch that moves in a metal circular waveguide having the deeply corrugated and smooth areas was investigated. The analytical solution was obtained using the equivalent boundary conditions for the case when the wavelengths are much greater than the structure period but the depth of the corrugated structure is of the same order as the wavelength under consideration. Two cases are considered: the bunch is flying out of the corrugated domain into a smooth one, and, inversely, the bunch is flying into the smooth domain out of the corrugated area of waveguide. The study was mainly focused on the analysis of the discrete part of the field.

The radiation in the regular infinite corrugated waveguide was also investigated. The frequency of radiated waves was found numerically and analytically for different charge velosities. It was shown that radiation can be both both multi- and quasi single-mode. It is generated at any velocity of the bunch motion.

In the first case when the bunch flies out of the corrugated area into the smooth one, several essential modes of the field discrete part can be transmitted through the boundary. The other modes are totally reflected off the boundary. The field discrete part can be the main part of wave field in the smooth area of the waveguide under some condition. The fronts of the modes of the field discrete part propagate with the group velocities.

In the second case when the bunch flies out of a smooth area into the corrugated one, the field discrete part is formed in some part of the corrugated region behind the bunch, and the rear front of this region moves with the group velocity in the same direction as the bunch.

The considered phenomenon can be prospective for generation of the radiation in the GHz and THz regions.

**Acknowledgments**
This research was supported by the Russian Science Foundation (Grant No. 18-72-10137).